\documentclass[final,3p,times,twocolumn]{elsarticle}

\usepackage{amssymb}
\usepackage{url}

\biboptions{sort&compress}

\journal{Physics Letters B}

\begin{document}

\begin{frontmatter}



\title{Production of neutron-rich nuclei in fragmentation reactions of $^{132}$Sn projectiles at relativistic energies.}


\author[USC]{D. P\'erez-Loureiro\corref{cor1}}
\cortext[cor1]{Corresponding author.}
\ead{david.loureiro@usc.es}
\author[USC]{J. Benlliure}
\author[USC]{H. \'Alvarez-Pol}
\author[CENBG]{B. Blank}
\author[USC]{E. Casarejos\fnref{label1}}
\fntext[label1]{Present address: University of Vigo}
\author[USC]{D. Dragosavac}
\author[GSI]{V. F\"ohr}
\author[USC]{M. Gasc\'on}
\fntext[label2]{Present address: Stanford University}
\author[UW]{W. Gawlikowicz}
\author[YALE]{A. Heinz\fnref{label2}}
\fntext[label3]{Present address: Chalmers University}
\author[UH]{K. Helariutta}
\author[GSI]{A. Keli\'c}
\author[GSI]{S. Luki\'c\fnref{label3}}
\fntext[label4]{Present address: University of Karlsruhe}
\author[GSI]{F. Montes\fnref{label4}}
\fntext[label5]{Present address: NSCL/MSU}
\author[UW]{L. Pie\'nkowski}
\author[GSI]{K-H. Schmidt}
\author[GSI]{M. Staniou}
\author[VINCA]{K. Suboti\'c}
\author[GSI]{K. S\"ummerer}
\author[CEA]{J. Taieb}
\author[UW]{A. Trzci\'nska}

\address[USC]{Universidade de Santiago de Compostela, E-15782 Spain}
\address[CENBG]{Centre d'Etudes Nucleaires, F-33175 Bordeaux-Gradignan Cedex, France}
\address[GSI]{GSI Helmholtzzentrum  f\"ur Schwerionenforschung, D-64291 Darmstadt, Germany}
\address[UW]{Heavy Ion Laboratory, University of Warsaw, PL-02-093 Warsaw, Poland}
\address[YALE]{A. W. Wright Nuclear Structure Laboratory, Yale University, New Haven, CT, 06520  USA}
\address[UH]{University of Helsinki, FI-00014 Helsinki, Finland}
\address[VINCA]{Institute of Nuclear Sciences Vin\v{c}a, University of Belgrade, 11001 Belgrade, Serbia}
\address[CEA]{CEA DAM, DPTA/SPN, BP. 12, 91680 Bruy\`eres-le-Ch\^atel, France}

\begin{abstract}
The fragmentation of neutron-rich $^{132}$Sn nuclei produced in the fission of $^{238}$U projectiles
 at 950 MeV/u has been investigated at the FRagment Separator (FRS) at GSI. This work represents the first investigation of fragmentation of medium-mass radioactive projectiles with a large neutron excess. The measured production cross sections of the residual nuclei are relevant for the possible use of a two-stage reaction scheme (fission+fragmentation) for the production of extremely neutron-rich medium-mass nuclei in future rare-ion-beam facilities. Moreover,  the new data will provide a better understanding of the ``memory'' effect in fragmentation reactions.
\end{abstract}

\end{frontmatter}


\section{Introduction}
\label{sec:intro}

Fragmentation reactions are widely used in rare-ion-beam facilities based on the in-flight separation technique for producing nuclei far from stability \cite{Lewitowicz:1994,Schneider:1994}. Indeed, this reaction mechanism contributed significantly to enlarge the present limits of the chart of nuclides reaching the proton drip-line up to mercury \cite{benlliure:2004}. The next generation of rare-ion-beam facilities \cite{RIBF,FAIR,FRIB} will certainly provide major advances in the neutron-rich sector. 

In this context, fragmentation has also proven to be the optimum option for producing heavy \cite{Alvarez:2010,Benlliure:2007} and light \cite{Baumann:2007} neutron-rich nuclei while medium-mass neutron-rich nuclei are mostly synthetized using fission reactions \cite{Bernas:1994,Bernas:1997,Ohnishi:2010}. Because of the large cross sections of neutron- or gamma-induced fission of actinides, and the neutron-excess of the fission residues, this reaction mechanism is known to produce high yields of medium-mass nuclei with a neutron excess close to that of the double magic $^{132}$Sn. Then, the yields decrease dramatically for more exotic nuclei \cite{Armbruster:2004,Benlliure:2008}. Thus, it is rather difficult to produce extremely neutron-rich nuclei along the N=82 shell or completely covering the region of nuclei involved in the stellar nucleosynthesis r process close to this shell.

From a technical point of view, ISOL facilities produce the largest in-target yields of fission products \cite{Shergur:2002}, however, the refractory nature of many of them results in rather poor extraction efficiencies, in particular for those nuclei with the largest neutron excess and consequently with the shortest half lives \cite{Lukic:2006}.  

In order to overcome these limitations, it was proposed some years ago to use a two-step reaction scheme for producing extremely neutron-rich medium-mass fragments \cite{Helariutta:2003}. According to this idea, one could profit from the large production yields of some fission fragments in an ISOL facility and fragment them after re-acceleration. In this way one could not only produce even more neutron-rich nuclei, but also cover the region of refractory elements poorly populated when using direct ISOL methods. The fragmentation of unstable projectiles has only been investigated with relatively light projectiles not too far from stability \cite{Lukyanov:2009}. Moreover, codes describing the production cross sections of residual nuclei in fragmentation reactions, such us COFRA \cite{Benlliure:1999} or EPAX \cite{Suemmerer:2000} provide rather discrepant predictions for the fragmentation of $^{132}$Sn \cite{Helariutta:2003}. These discrepancies could be due to strong memory effects that one could expect when investigating projectiles with a large neutron excess that are not properly implemented in these codes.

In this Letter we present the first measurement of fragmentation of fission fragments, in particular $^{132}$Sn. These data are relevant not only for solving the discrepancies between different fragmentation codes and for validating the two-reaction scheme for the production of extremely neutron-rich medium-mass fragments, but also for investigating the memory effect in fragmentation reactions.

\section{The experiment}

The experiment was performed at the GSI facilities in Darmstadt, Germany. The SIS18 synchrotron delivered a $^{238}$U beam at 950 MeV/u with an average intensity of 10$^8$ particles per second. The FRagment Separator (FRS) \cite{Geissel:1992} was used to isotopically identify the nuclei produced in the reactions investigated in this work. The FRS is a zero-degree magnetic spectrometer composed of two symmetric stages. Each stage is composed of two dipoles and sets of quadrupoles defining an optical system with four focal planes, one behind each dipole magnet.

In this particular experiment, the FRS was equipped with two reaction targets. A  650 mg/cm$^2$ lead target was installed at the entrance of the FRS enhancing the fission of the $^{238}$U projectiles through the Coulomb interaction. A  2.6 g/cm$^2$ beryllium target was placed at the second focal plane of the spectrometer to induce the fragmentation of the fission fragments produced in the first target. In this configuration, the two sections of the FRS were used as two independent spectrometers. The first section provided the selection and identification of fission fragments produced in the reactions induced by the $^{238}$U projectiles in the lead target, in particular $^{132}$Sn. The second section was used to identify the residual nuclei produced in the reactions induced by the fission fragments impinging onto the beryllium target. 

Both sections of the FRS were equipped with tracking, time-of-flight and energy-loss detectors. Time-projection chambers provided the position of the nuclei at the intermediate and final focal planes of the spectrometer used to determine their magnetic rigidity. Three plastic scintillators placed at the first, the second and the final image planes were used to obtain the velocities of the ions in the two sections. Finally, two ionization chambers located at the second and  final image plane, respectively, provided the identification in atomic number from the energy lost by the nuclei traversing these detectors. The combined measurement of the magnetic rigidity (B$\rho$), velocity ($\beta \gamma$) obtained from time of flight and atomic number of fully stripped ions ($Z$) made it possible to unambiguously identify the nuclear species under investigation through the expression \(B\rho=\gamma \beta m_o c/Ze\). A detailed description of the experimental setup can be found in \cite{Cortina:2003}.

The main challenge of this experiment was the isotopic identification of fission fragments with mass numbers around 130 moving with relativistic velocities using only the first section of the FRS. Because of the large angular spreading of fission fragments an accurate tracking of their trajectories was required. Due to the presence of the primary beam in front of the first dipole, the time of flight could only be measured between the first and the second focal planes with a limited flight path of around 17 meters. Moreover, the large fraction of fission fragments transmitted through the first dipole required detectors supporting huge counting rates at the first image plane. These difficulties were overcome by using time projection chambers for tracking and plastic scintillators equiped with H2431-50MOD Hamamatsu photomultipliers with a booster base, providing an intrinsic time resolution around 70 ps (FWHM) and accepting counting rates larger than 100 kHz. 

 \begin{figure}[ht]
\includegraphics[width=.45\textwidth]{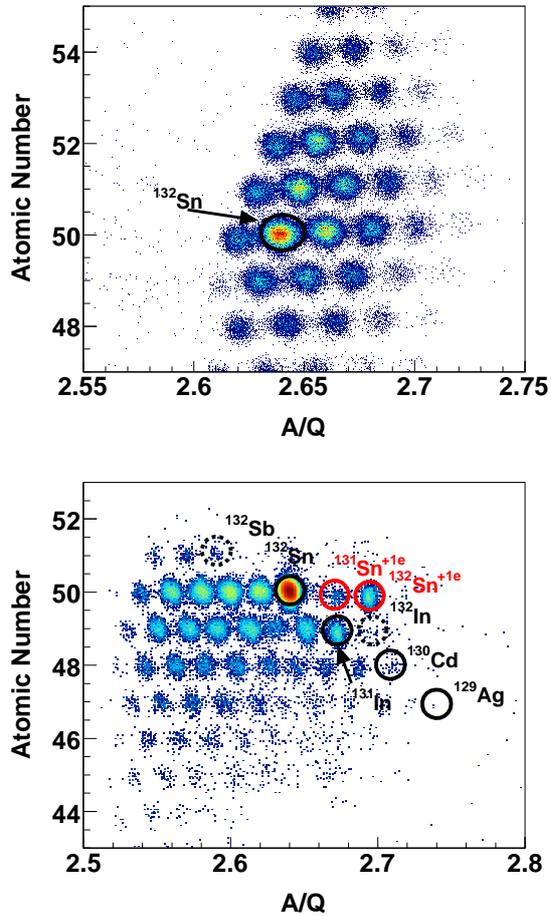}
\caption{Upper panel: Part of the identification matrix of fission residues transmitted to the intermediate focal plane of the Fragment Separator. $^{132}$Sn nuclei are marked by the black circle. Lower panel: Identification matrix of the residual nuclei produced in the fragmentation of $^{132}$Sn.}
\label{fig_1}
\end{figure}

In this experiment, the first section of the FRS was tuned to maximize the transmission of $^{132}$Sn, while several settings of the second section were required in order to cover the most neutron-rich nuclei produced in the fragmentation of $^{132}$Sn with beryllium. Figure \ref{fig_1} displays the two identification matrices obtained in sorting the data measured in this experiment. The upper plot corresponds to the identification plot of fission fragments transmitted until the intermediate focal plane of the spectrometer. As can be seen, despite the limited flight path used for the determination of the time of flight of these nuclei and the large counting rates ($>$ 100 kHz) we were able to obtain an unambiguous separation of all fission fragments transmitted around $^{132}$Sn.

The lower plot in Fig. \ref{fig_1} was obtained combining different tunings of the second stage of the FRS optimizing the transmission of $^{132}$Sb, $^{132}$Sn, $^{131}$In, $^{123}$Ag, $^{126}$Ag and $^{128}$Pd selecting only those events for which a $^{132}$Sn nucleus was identified at the intermediate focal plane of the spectrometer. As can be seen in the figure, these tunings allowed us to cover a large fraction of the neutron-rich nuclei obtained in the fragmentation of $^{132}$Sn. In particular, we were able to identify the most neutron-rich isotopes of In, Cd and Ag  that can be produced in this reaction, corresponding to the one-proton ($^{131}$In), two-proton ($^{130}$Cd) and three-proton ($^{129}$Ag) removal channels. These reaction channels are extremely important for the understanding of the fragmentation process since in this case only protons are abraded from the projectile, and the excitation energy gained by the resulting residual nuclei remains below the neutron evaporation threshold. In this figure one can also identify the production of $^{130,131,132}$Sb and $^{132}$In in (n,p) \cite{Kelic:2004} and (p,n) \cite{Morales:2011} charge exchange reactions.

\section{Measured cross sections}

The excellent resolving power shown in Fig. \ref{fig_1}, both for the fission fragments and for the fragmentation residues, allowed us to determine for the first time accurate values for the production cross sections of residual nuclei produced in fragmentation reactions induced by radioactive projectiles with a large neutron excess. In order to determine the corresponding production cross sections, the yields of fragmentation residues measured at the final focal plane of the FRS (different spots in the lower panel of Figure \ref{fig_1}) were corrected for optical transmission effects between the intermediate and final focal planes of the spectrometer. These corrected yields were then normalized to the number of $^{132}$Sn projectiles arriving at the intermediate focal plane (spots in the upper panel of Figure \ref{fig_1}) and to the number of atoms per surface unit of the target in order to obtain the production cross sections.

\begin{figure}[ht]
\centering
\includegraphics[width=.45\textwidth]{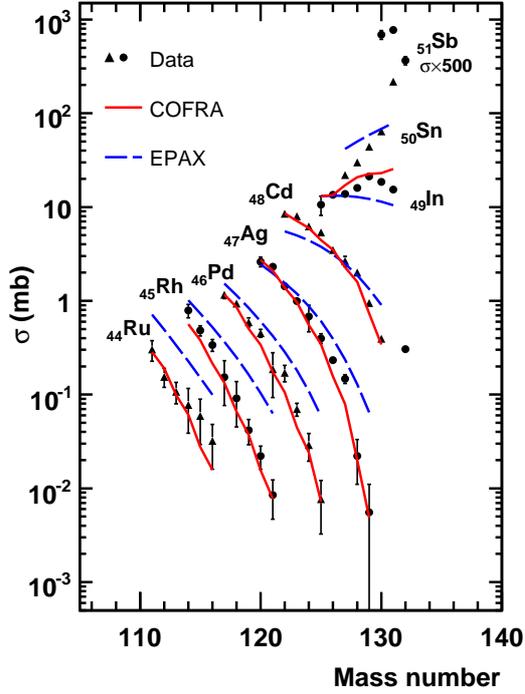}
\caption{Isotopic distributions of the production cross sections of residual nuclei measured in the fragmentation of $^{132}$Sn on beryllium. Error bars are shown if  bigger than symbols. The lines correspond to two different model calculations COFRA \cite{Benlliure:1999} (solid line) and EPAX \cite{Suemmerer:2000} (dashed line).}
\label{fig_2}
\end{figure}

In Figure \ref{fig_2} we present the isotopic distributions of the most neutron-rich residual nuclei produced in the fragmentation of $^{132}$Sn on beryllium at 950 AMeV. As can be seen, we were able to identify and determine the production cross section of 55 of these fragmentation residues of elements ranging from Sn (Z=50) to Ru (Z=44). The lowest cross sections measured in this experiment are of the order of a few $\mu$b. This limit is explained  by the limited production of fission fragments ($\approx$10$^3$ $^{132}$Sn s$^{-1}$), as compared to similar stable projectiles such as $^{136}$Xe \cite{Benlliure:2008}.

\section{Model calculations}

The data obtained in this work clearly demonstrate that the fragmentation of fission fragments leads to a sizeable production of medium-mass fragments with large values of neutron excess and, therefore, confirm the validity of the two-step reaction scheme to extend the limits of the chart of nuclides in this region. However, due to the limited intensity of fission fragments that could be fragmented in this experiment, a quantitative conclusion on the production of neutron-rich nuclei in this region, e.g. reachable with a high-intensity $^{132}$Sn beam produced in a future ISOL facility, requires the extrapolation of the present results using reliable model calculations.  

In Fig. \ref{fig_2} we compare the measured cross sections with the predictions obtained with two of the most widely used calculation codes describing the production cross sections of nuclei far from stability, COFRA \cite{Benlliure:1999} (solid line) and EPAX \cite{Suemmerer:2000} (dashed line). As can be clearly seen, COFRA provides a rather accurate description of the measured cross sections while EPAX overestimate the production of the most neutron-rich fragments. Similar conclusions were obtained investigating the fragmentation of stable projectiles such as $^{136}$Xe \cite{Benlliure:2008}, $^{208}$Pb \cite{Benlliure:2007} and $^{238}$U \cite{Alvarez:2010}. A possible explanation of these results is given in the following.

COFRA is a simplified version of the abrasion-ablation reaction model by Gaimard and Schmidt \cite{Gaimard:1991}. COFRA includes a full description of the abrasion process where the number of nucleons removed from projectile/target is described according to the geometrical approach \cite{Oliveira:1979} and the excitation energy gained in this process is obtained from the energy distribution of the holes produced by the abraded nucleons \cite{Schmidt:1993}. The second stage of the reaction is then described under the assumption that residual nuclei with a large neutron excess will exclusively de-excitate by neutron evaporation. This assumption allows to simplify the  statistical evaporation using an analytical formulation that drastically reduces the computation time \cite{Benlliure:1999}. In this approach, the probability for neutron evaporation, and thus, the cross section for the final residual fragments is governed by the excitation-energy distribution and the binding energies of the neutrons in the pre-fragments. The good agreement of the data shown in Fig. \ref{fig_2} in addition to previous data obtained with stable projectiles \cite{Benlliure:1999,Benlliure:2008,Alvarez:2010}, demonstrate the appropiate description of abrasion process in the model and the clear dependence of the final production cross sections on the binding energies of the nuclei of interest.

While in COFRA the isospin dependence of the fragmentation process is mostly ruled by the binding energies of the involved nuclei, in the case of the EPAX formula this dependence is parameterized in terms of the so called ``memory'' effect \cite{Summerer:1990}. This ``memory'' takes into account the difference in neutron excess between the projectile and the corresponding nuclei at the evaporation corridor \cite{Porile:1979}. Figure \ref{fig_2} (dashed line) clearly shows that the parameterization of this effect in EPAX is not optimal. The most probable reason is that the data used to fix this parameter in EPAX ($^{208}$Pb, $^{129}$Xe and $^{86}$Kr) do not cover so large a range in neutron excess as the ones obtained in this work.   

\begin{figure}[b]
\centering
\includegraphics[width=.45\textwidth]{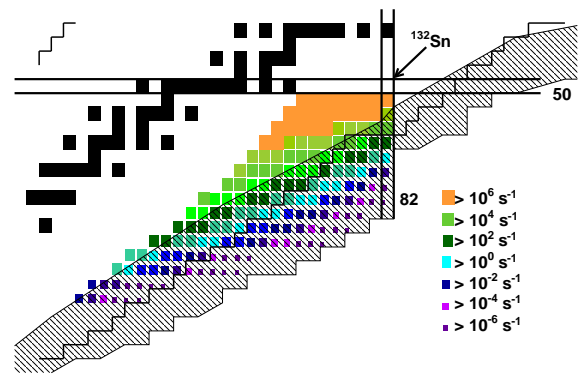}
\caption{Estimated production rates of extremely neutron-rich medium-mass nuclei in fragmentation reactions induced by a 1.4 10$^{11}$ pps $^{132}$Sn beam impinging onto a 1 g/cm$^2$ beryllium target obtained with the code COFRA. }
\label{fig_3}
\end{figure}

From the previous analysis one can easily conclude that the COFRA code is well suited for predicting the production of neutron-rich residual nuclei in fragmentation reactions. Therefore, we have used this code to extrapolate the results obtained in the present work. In Fig. \ref{fig_3} we display the predicted production rates of neutron-rich residual fragments from the fragmentation of a $^{132}$Sn beam with an intensity of 1.4 10$^{11}$ ions s$^{-1}$ impinging onto a 1 g/cm$^2$ thick beryllium target. This scenario corresponds to the two-stage reaction scheme proposed for the future EURISOL facility \cite{EURISOL} where the high-power target station could produce up to 10$^{15}$ neutron-induced fissions of $^{238}$U per second. Taking into account an extraction efficiency from the ISOL stage for $^{132}$Sn of 42\% \cite{Lukic:2006} and a total post-acceleration efficiency of 10\% \cite{EURISOL} one obtains the mentioned intensity for a $^{132}$Sn beam accelerated at few hundred MeVs per nucleon. The results indicate that the proposed production scheme would not only overcome the limited efficiency of ISOL facilities for the extraction of isotopes of refractory elements between Zr and Rh, but also would produce a large variety of neutron-rich nuclei in this region of the chart of nuclides. Indeed, according to these estimates, the fragmentation of such a $^{132}$Sn would almost entirely populate the region of nuclei around the A$\approx$130 waiting point of the r process stellar nucleosynthesis.

\section{Conclusions}
In this Letter we have presented the first measurements on fragmentation reactions induced by medium-mass neutron-rich secondary beams produced in fission reactions of a $^{238}$U primary beam. In particular, we have determined the production cross sections of 55 residual nuclei produced in the fragmentation of $^{132}$Sn with a beryllium target at an energy around 950 A MeV. The accuracy of these measurement rely on the clear separation and identification of the residual nuclei produced in both reactions, fission and fragmentation, obtained with an improved detection setup at the FRS spectrometer at GSI.

The measured cross sections were used to benchmark the most widely used model calculations for predicting the production cross sections of residual nuclei in fragmentation reactions, COFRA and EPAX. The good agreement of the present data with COFRA indicates the appropriate modelling of the abrasion process in this code and confirms that the production cross sections of the most neutron-rich fragmentation residues are governed both by neutron evaporation and  the neutron binding energies of these nuclei. The overestimation of the cross sections with the EPAX formula was explained as due to a deficient description of the ``memory'' effect. The new data could be used to improve the predictions of EPAX. 

Finally, the code COFRA was used to extrapolate the present results in order to investigate the possibility of using a two step reaction scheme (fission+fragmentation) to overcome the limited production of refractory exotic nuclei in ISOL facilities but also to estimate the production of extremely neutron-rich nuclei of medium mass.

\section*{Acknowlegments}
This work was partially funded by the Spanish Ministry for Science and Innovation under grant FPA2007-6252, the programme ``Ingenio 2010, Consolider CPAN'', the regional government of Galicia under grant  No. PGIDT00PXI20606PM and EC under the  EURISOL-DS contract No. 515768 RIDS.





\bibliographystyle{model1-num-names}
\bibliography{bibliography}






\end{document}